# The interaction of frictional slip and adhesion for a stiff sphere on a compliant substrate

R.M. McMeeking[1,2,3,4], M. Ciavarella[5,6], G. Cricrì[7] and K.-S. Kim[8]

[1]Materials Department, University of California, Santa Barbara CA 93106
[2]Department of Mechanical Engineering, University of California, Santa Barbara CA 93106
[3]School of Engineering, Aberdeen University, King's College, Aberdeen AB24 3UE, Scotland
[4]INM – Leibniz Institute for New Materials, Campus D2 2, 66123 Saarbrücken, Germany
[5]Politecnico di BARI, DMMM Dept., Via Edoardo Orabona 4, 70126 Bari, Italy
[6]Hamburg University of Technology, Department of Mechanical Engineering, Am Schwarzenberg-Campus 1, 21073 Hamburg, Germany
[7]Università di Napoli Federico II, DII Dept., Piazzale Vicenzo Tecchio 80, 80125 Napoli, Italy
[8]School of Engineering, Brown University, Providence RI 02912

Dedicated to John Hutchinson on the occasion of his 80th birthday.

**Abstract**
How friction affects adhesion is addressed. The problem is considered in the context of a very stiff sphere adhering to a compliant, isotropic, linear elastic substrate, and experiencing adhesion and frictional slip relative to each other. The adhesion is considered to be driven by very large attractive tractions between the sphere and the substrate that can act only at very small distances between them. As a consequence, the adhesion behavior can be represented by the Johnson-Kendall-Roberts model, and this is assumed to prevail also when frictional slip is occurring. Frictional slip is considered to be resisted by a uniform, constant shear traction at the slipping interface, a model that is considered to be valid for small asperities and for compliant elastomers in contact with stiff material. A model for the interaction of friction and adhesion, known to agree with some experimental data, is utilized. This model is due to Johnson, and its adhesion-friction interaction is assumed to stem, upon shrinkage of the contact area, from a postulated reversible energy release associated with frictional slip. This behavior is considered to arise from surface microstructures generated or eliminated by frictional slip, where these microstructures store some elastic strain energy in a reversible manner. The associated reversible energy release rate is derived from the energy exchanges that occur in the system. The Johnson model, and an asymptotic analysis of it for small amounts of frictional slip, is shown to be consistent with the reversible energy release rate that we identify.

**Introduction**
There is significant interest in the question of how frictional slip affects adhesion between solid objects. An exemplar for this topic is a stiff sphere interacting with a compliant, flat substrate when the radius of the contact is small compared to the radius of the sphere, though the converse of a stiff substrate interacting with a compliant sphere behaves, in principle, in an equivalent manner. It is also possible to analyze problems in which both the sphere and the substrate are compliant, but with added complexity so that the issues cannot be elucidated as clearly as when one component is very stiff. Thus, we will address the issues germane to the interaction of adhesion and friction by consideration of the stiff sphere and the compliant substrate.

A reason for pursuing this topic in a festschrift dedicated to John Hutchinson is that there are aspects of this area that are reminiscent of mixed mode fracture mechanics, a subject in which John did much pioneering work. Therefore, this paper serves to remind RMM of the happy days of some time ago when John and RMM worked within a broader collaborative group addressing research on mixed mode fracture in the context of advanced composite materials. KSK also has fond memories of learning mixed-mode fracture mechanics from John to solve a thin film detachment problem, and is very pleased to contribute to this paper in his honor. We add that the co-authors of this paper are thrilled that John is ranked number one in mechanical engineering in the world according to recent findings by Ioannidis *et al*. (*PLOS Biology*, 17(8): e3000384), and note that it comes as no surprise to us.

We should comment that much of what is to be presented in this paper covers well-trodden ground and no profoundly new results are presented. Prior publications that we will rely heavily on are by Johnson [1] and Kim, McMeeking and Johnson [2]. The former presented a model for the interaction of friction and adhesion that relied on concepts of mixed mode fracture mechanics. The latter elucidated issues concerning the meaning of energy release rates when frictional slip is present in the interaction of an adherent sphere with a flat substrate. By doing so, Kim *et al*. [2] put the model of Johnson [1] into context, proving it to be meaningful over a range of conditions that can be encountered in adhesion and friction, including large degrees of slip. However, the current paper serves to clarify and illustrate further some ideas from [1] & [2] that apply to how frictional slip can influence adhesion.

We consider the geometry illustrated in Fig. 1, where a rigid sphere of radius $R$ is adhered, through a contact circle of radius $a$, to a compliant, linear elastic, isotropic substrate. The substrate has elastic modulus $E$ and Poisson's ratio $\nu$. The sphere is subjected to an applied load $P$ in compression and another, $T$, parallel to the substrate surface, with the sphere experiencing a displacement $\Delta$ towards the substrate and $D$ parallel to it. Note that the force $T$ parallel to the substrate is shown in Fig. 1 applied to the sphere at its lowest point; the reason for this will become clear below. However, the reader is reminded that elementary considerations of statics for a rigid body allows us to apply a combination of moments and forces anywhere to the sphere that is statically equivalent to the force, $T$, shown in Fig. 1, thereby avoiding the difficulty of actually applying the force at the bottom of the sphere.

It is assumed that $a \ll R$ so that Hertzian contact [3], JKR adhesion theory [4], the Maugis-Dugdale model [5] and the shearing model of Keer and Goodman [6] can be exploited in the appropriate regimes. For the JKR and Maugis-Dugdale models we take the adhesion energy in the absence of motion and loading parallel to the substrate to be $w_o$. It is equal to $\gamma_1 + \gamma_2 - \gamma_{12}$, where $\gamma_1$ and $\gamma_2$ are the effective specific surface energies of, respectively, the sphere and the substrate when exposed to air, and $\gamma_{12}$ is the effective specific energy of the interface between the sphere and the substrate when touching. When the material of the sphere slips relative to the substrate it is assumed that the slipped surface experiences a constant, uniform shear traction, $\tau_o$. As discussed by Johnson [1], a friction model with uniform, constant shear traction is valid for a single asperity in a rough surface. A friction model with uniform, constant shear traction is also appropriate for a compliant elastomer in contact with a stiff material [7].

When there is adhesion without tangential motion or loading, and adhesive tractions are very large acting at very small distances, the JKR [4] solution prevails and we have

$$P = 2aE^* \left(\Delta - \frac{a^2}{3R}\right) \tag{1}$$

where, in this case, the reduced modulus, $E^*$, is given by

$$E^* = \frac{E}{1-\nu^2} \tag{2}$$

a result that arises because the sphere is rigid. The strain energy in the substrate is

$$\mathcal{E}_\Delta = E^* \left(a\Delta^2 - \frac{2a^3\Delta}{3R} + \frac{a^5}{5R^2}\right) \tag{3}$$

The strain energy release rate, upon reduction of the area of contact, $A = \pi a^2$, is

$$\mathcal{G}^{SE} = \frac{d\mathcal{E}_\Delta}{dA} - P\frac{d\Delta}{dA} = \frac{1}{2\pi a}\frac{\partial \mathcal{E}_\Delta}{\partial a} + \left(\frac{\partial \mathcal{E}_\Delta}{\partial \Delta} - P\right)\frac{d\Delta}{dA} = \frac{1}{2\pi a}\frac{\partial \mathcal{E}_\Delta}{\partial a} = \frac{E^*}{2\pi a}\left(\frac{a^2}{R} - \Delta\right)^2 \tag{4}$$

Note that in evaluating Eq. (4) there is no need to specify that any variables, such as $P$ or $\Delta$, remain constant. Similarly, if we obtain the complementary strain energy, *i.e.* the potential energy, by a Legendre transformation

$$\Phi_P = \mathcal{E}_\Delta - P\Delta = E^* a \left[\frac{4a^4}{45R^2} - \frac{Pa}{3E^*R} - \left(\frac{P}{2E^*a}\right)^2\right] \tag{5}$$

we deduce the strain energy release rate as

$$\mathcal{G}^{SE} = \frac{d\Phi_P}{dA} + \Delta\frac{dP}{dA} = \frac{1}{2\pi a}\frac{\partial \Phi_P}{\partial a} + \left(\frac{\partial \Phi_P}{\partial P} + \Delta\right)\frac{dP}{dA} = \frac{1}{2\pi a}\frac{\partial \Phi_P}{\partial a} = \frac{E^*}{2\pi a}\left(\frac{2a^2}{3R} - \frac{P}{2E^*a}\right)^2 \tag{6}$$

Again, there is no need to specify that any variables, such as $P$ or $\Delta$, remain constant. Use of Eq. (1) confirms that Eq. (4) & (6) give the same result.

In the JKR setting, the reversible energy release rate, $\mathcal{G}^{re}$, in the absence of frictional slip, is equal to $\mathcal{G}^{SE}$ and $\mathcal{G}^{re} = w_o$. We then obtain the equilibrium solution [4]

$$\Delta = \frac{a^2}{R} - \sqrt{\frac{2\pi a w_o}{E^*}} \tag{7}$$

$$P = \frac{4E^* a^3}{3R} - \sqrt{8\pi a^3 E^* w_o} \tag{8}$$

Although we will allude to the Maugis-Dugdale [5] regime, where peak interaction tractions are smaller than for the JKR regime but the interaction distances are larger, we will not provide any details. For them, the reader is directed to Maugis' [5] paper and to a summary of it in Kim *et al*. [2].

To address a contact that is experiencing shear motion parallel to the substrate as well as adhesion, we first make some general observations regarding energy balance. We write this in the form

$$\frac{dQ}{dt} = P\frac{d\Delta}{dt} + T\frac{dD}{dt} - \frac{d\mathcal{E}_\Delta}{dt} - \frac{d\mathcal{E}_D}{dt} - \frac{d\mathcal{E}_S}{dt} + \frac{d(w_o \pi a^2)}{dt} \qquad (9)$$

where $dQ/dt$ is the rate at which heat is generated, $\mathcal{E}_D$ is the elastic strain energy in the substrate due to the shearing loads applied to it due the force $T$ and the motion $D$, and $\mathcal{E}_S$ is any reversible work stored in the interface due to slip. We will return to $\mathcal{E}_S$ later to clarify its meaning, but the important point for the time being is that it is reversible so that some work added to the interface by slip can be released again. Note that we assume that $\mathcal{E}_D$ is decoupled from normal loading and $\mathcal{E}_\Delta$ is decoupled from shear loading, a situation we can rely on because of the isotropy of the substrate and due to freedom in the manner by which the sphere's motion can be controlled. Specifically, because we apply $T$ at the base of the sphere we do not apply any moment to the contact zone.

We first consider the situation where adhesion is unaffected by slip. As a result

$$P\frac{d\Delta}{dt} = \frac{d\mathcal{E}_\Delta}{dt} - \frac{d(w_o \pi a^2)}{dt} \qquad (10)$$

consistent with the JKR result. In addition, $\mathcal{E}_S = 0$, for if it were not, the reversible change in energy associated with it would be available to influence adhesion. As a consequence, when slip has no influence on adhesion, Eq. (9) becomes

$$\frac{dQ}{dt} = T\frac{dD}{dt} - \frac{d\mathcal{E}_D}{dt} \qquad (11)$$

This equation states that any work done by $T$ that is not absorbed by the strain energy of the substrate causes the generation of heat. Or, in other words, due to friction at the interface, any slip that occurs there will create heat. We note that this deduction is rather general, and does not depend on the specifics of the model for frictional slip. We note that a converse of this observation is that the only destinations for the energy made available by any reduction of $\mathcal{E}_D$ is (1) the generation of heat, and (2) the reduction of the rate of increase of the effective value of $D$ that characterizes the aggregate amount of slip that the sphere has experienced. This observation is consistent with the starting assumption that slip cannot affect adhesion, since the destination of $-d\mathcal{E}_D/dt$ would have to be one of the terms in Eq. (10) if slip were going to affect adhesion through the release of energy stored in $\mathcal{E}_D$.

Now consider the case where slip *does* affect adhesion. In that case, we cannot separate out the JKR terms embedded in Eq. (9) and we allow $\mathcal{E}_S$ to be nonzero in general. Thus, we must utilize Eq. (9) as stated when slip has an effect on adhesion.

**The Slip Model**
Our model for friction and slip is due to Keer and Goodman [6], rederived by Savkoor [8] in his doctoral thesis, and summarized also by Johnson [1] and Kim *et al*. [2]. As we shall see below, this model has some limitations, but it serves as a useful formulation as it has explicit analytical expressions. Consistent with Johnson [1] and Kim *et al*. [2], we take the contact interface having nonzero frictional tractions to be identical to that having normal tractions. In addition, we assume that there has been no slip within a zone such that $r \leq b$ and that for $b \leq r \leq a$ there is a uniform shear traction $\tau_o$ acting on the sphere to oppose its sliding (see Fig. 1 & 2). In these definitions, $(r, \theta, z)$ is an orthonormal polar coordinate system with origin at the center of the contact circle and with the $z$ direction orthogonal to the surface of the substrate (see Fig. 2). Note that in the slipping zone the shear traction is emphasized to be uniform, constant and everywhere acting in the same direction. In addition, we assume that the displacement, $D$, of the sphere parallel to the substrate surface is monotonically increasing. In such circumstances [1, 2, 6, 8]

$$D = \frac{(2-\nu)\tau_o}{(1-\nu)E^*}\sqrt{a^2 - b^2} \qquad\qquad D \leq \frac{(2-\nu)\tau_o a}{(1-\nu)E^*} \qquad (12)$$

$$T = 2\tau_o \left(b\sqrt{a^2 - b^2} + a^2 \cos^{-1}\frac{b}{a}\right) \qquad (13)$$

As noted by Kim *et al*. [2], when $D$ is increased monotonically from zero at constant $a$ the accumulated work done by $T$ is

$$U_D = \frac{2(2-\nu)\tau_o^2}{(1-\nu)E^*}\left(a^2\sqrt{a^2 - b^2} \cos^{-1}\frac{b}{a} + a^2 b - \frac{b^3 + 2a^3}{3}\right) \qquad D \leq \frac{(2-\nu)\tau_o a}{(1-\nu)E^*} \qquad (14)$$

The situation described by Eq. (12) to (14) is defined as limited slip as the region within $r \leq b$ has not experienced any relative sliding.

When $D$ continues to be increased monotonically beyond $(2 - \nu)\tau_o a/[(1 - \nu)E^*]$, the entire interface is slipping, $b = 0$, the shear traction magnitude everywhere at the interface is $\tau_o$, and the applied force parallel to the substrate becomes $T = \tau_o \pi a^2$. This situation is termed gross slip. In addition, we note that for monotonic motion at constant $a$

$$U_D = \frac{2(2-\nu)\tau_o^2 a^3}{(1-\nu)E^*}\left(\frac{\pi}{2} - \frac{2}{3}\right) + \tau_o \pi a^2 \left[D - \frac{(2-\nu)\tau_o a}{(1-\nu)E^*}\right] \qquad D \geq \frac{(2-\nu)\tau_o a}{(1-\nu)E^*} \qquad (15)$$

where the 1st term on the right hand side is the work done by $T$ while $b$, commencing at $a$, diminishes to zero, and the 2nd term is the work done by $T$ after gross slip has begun, in which case $b = 0$. We note that when gross slip is taking place

$$T = \frac{\partial U_D(D,a)}{\partial D} = \tau_o \pi a^2 \tag{16}$$

as noted above, and reflecting the fact that the entire contact is sliding. In addition, subject to monotonic increase, when Eq. (15) prevails $D$ is arbitrary and is thus an independent variable.

In regard to $U_D$, some of it is stored as strain energy of distortion of the substrate, some is dissipated as heat due to frictional sliding, and some may be stored or unstored reversibly in surface microstructures at the nanoscopic scale [1, 2]. The exact mix of this partition of energy is difficult to determine without nanoscopic experiments that would be challenging to undertake, or complex atomistic calculations of the possible processes involved. Therefore, we do not attempt to specify the status of the energy in Eq. (14) & (15), other than to observe that the stored strain energy, including that injected into or removed from surface microstructures, may play a role that affects both adhesion and friction. We observe that the total strain energy in the substrate is $\mathcal{E}_\Delta + \mathcal{E}_D$, and we characterize the energy stored or unstored reversibly in surface microstructures as $\mathcal{E}_S$, thereby giving it a meaning.

During monotonic, limited slip, the $x$ component of displacement within the contact circle at the surface of the substrate is

$$u_x = D \qquad\qquad 0 \leq r \leq b \tag{17a}$$

$$u_x = D - \frac{2\tau_o}{\pi(1-v)E^*} \int_b^r \frac{\rho}{\sqrt{a^2-\rho^2}} \left[ (2-v)\cos^{-1}\frac{\rho}{r} - \frac{vt}{r^2}\sqrt{r^2-\rho^2}\cos 2\theta \right] dt \qquad b \leq r \leq a \tag{17b}$$

a result that is due to Keer and Goodman [6] but rederived by Savkoor [8]. Following a procedure used by Menga *et al.* [7], we compute the strain energy in the substrate as

$$\mathcal{E}_D = \frac{1}{2}\int_A \tau_x u_x dA = \frac{1}{2}TD - \frac{2(2-v)\tau_o^2}{(1-v)E^*}\int_b^a r \int_b^r \frac{\rho \cos^{-1}\frac{\rho}{r}}{\sqrt{a^2-\rho^2}} d\rho\, dr \tag{18}$$

where $\tau_x$ is the traction applied by the sphere to the substrate within the contact circle.

During gross slip, Menga *et al.* [7] have identified the tangential displacement of the substrate in the contact circle in the direction of motion as

$$u_x = \frac{2\tau_o a}{(1-v)\pi E^*}\left\{(2-v)L\left(\frac{r}{a}\right) + \frac{a^2 v}{3r^2}\left[\left(2-\frac{r^2}{a^2}\right)L\left(\frac{r}{a}\right) - 2\left(1-\frac{r^2}{a^2}\right)K\left(\frac{r}{a}\right)\right]\cos 2\theta\right\} \tag{19}$$

where $K(r/a)$ is the complete elliptic integral of the 1st kind and $L(r/a)$ is the complete elliptic integral of the 2nd kind. Note that in Eq. (19) we have generalized the result slightly compared to that of Menga *et al.* [7] to account for an arbitrary Poisson's ratio of the substrate. The result in Eq. (19) is obtained from analysis provided in Johnson [3], Menga *et al.* [7] and Savkoor [8], along with results to be found in Gradshteyn and Ryzhik [9]. In the absence of vertical constraint, the displacement within the contact circle in the vertical direction is, in gross slip,

$$u_z = -\frac{(1-2\nu)\tau_o x}{2(1-\nu)E^*} \tag{20}$$

with the positive sense of this displacement being towards the substrate. By defining $T$ to be applied at the bottom of the sphere, we avoid any constraint that will limit the displacement given by Eq. (20). Menga *et al.* [7] used the equivalent of Eq. (18) to compute the strain energy of distortion of the substrate during gross slip due to monotonic motion of the sphere as

$$\mathcal{E}_D = \frac{4(2-\nu)\tau_o^2 a^3}{3(1-\nu)E^*} \tag{21}$$

Again, we have generalized the result to allow for any value of Poisson's ratio.

Note that during gross slip the strain energy, $\mathcal{E}_D$, is independent of $D$ and depends only on the size of the contact.

**Energy Balance and the Slip Model**
We make use of Eq. (1) & (3) to obtain a modest simplification of Eq. (9) in the form

$$\frac{dQ}{dt} = T\frac{dD}{dt} - E^*\left(\frac{a^2}{R} - \Delta\right)^2 \frac{da}{dt} - \frac{d\mathcal{E}_D}{dt} - \frac{d\mathcal{E}_S}{dt} + 2\pi a w_o \frac{da}{dt} \tag{22}$$

*Case A: Slip has no effect on adhesion.*
As noted above the heat production rate in this case is given by Eq. (11). In limited slip, the extent of sliding that takes place during monotonic motion is given by

$$\delta = D - u_x = \frac{2\tau_o}{\pi(1-\nu)E^*}\int_b^r \frac{\rho}{\sqrt{a^2-\rho^2}}\left[(2-\nu)\cos^{-1}\frac{\rho}{r} - \frac{\nu\rho}{r^2}\sqrt{r^2-\rho^2}\cos 2\theta\right]d\rho$$
$$b \leq r \leq a \quad (23)$$

while it is zero within $0 \leq r \leq b$. The generation of heat obeys

$$\frac{dQ}{dt} = \tau_o \int_A \frac{d\delta}{dt} dA \tag{24}$$

which, in principle can be evaluated with use of Eq. (23), while the rate of change of the strain energy can be obtained, again in principle, from differentiation of Eq. (18). However, the integrals involved are challenging and we will defer the analysis until later in the paper.

For gross slip, we return to Eq. (22) and explore the consequence of using $\mathcal{E}_S = 0$ with the understanding that all slip work is dissipated as heat. In this case, from Eq. (19) we obtain

$$\delta = D - u_x = D - \frac{2\tau_o a}{(1-\nu)\pi E^*}\left\{(2-\nu)L\left(\frac{r}{a}\right) + \frac{a^2\nu}{3r^2}\left[\left(2-\frac{r^2}{a^2}\right)L\left(\frac{r}{a}\right) - 2\left(1-\frac{r^2}{a^2}\right)K\left(\frac{r}{a}\right)\right]\cos 2\theta\right\} \tag{25}$$

and consider the matter of evaluating the rate of slip to enable use of Eq. (24). Slip in Eq. (25) is such that

$$\delta = \delta(r, \theta, D, a) \tag{26}$$

where functional dependence on material parameters has been omitted. For a material point on the substrate surface each of the arguments in Eq. (26) can change with time, with all except $a$ definitely doing so during gross slip. Thus, for a material point in the contact surface we write

$$\frac{d\delta}{dt} = \left(\frac{\partial \delta}{\partial r}\frac{\partial r}{\partial D} + \frac{\partial \delta}{\partial \theta}\frac{\partial \theta}{\partial D} + \frac{\partial \delta}{\partial D}\right)\frac{dD}{dt} + \frac{\partial \delta}{\partial a}\frac{da}{dt} \tag{27}$$

Since Eq. (24) involves integrating over the entire slipping surface of the contact, symmetries in Eq. (25) will lead to cancellation of many of its nonzero terms upon integration. Therefore, there is no need, in general, to carry out a full evaluation of the terms in Eq. (27). However, it is insightful to consider the special case of $\nu = 0$ as this simplifies things greatly. In that case

$$\frac{d\delta}{dt} = \left\{1 + \frac{4\tau_o a}{\pi E^* r}\left[L\left(\frac{r}{a}\right) - K\left(\frac{r}{a}\right)\right]\cos\theta\right\}\frac{dD}{dt} - \frac{4\tau_o}{\pi E^*}K\left(\frac{r}{a}\right)\frac{da}{dt} \tag{28}$$

Now, assume that the system is experiencing gross slip in steady state, so that $da/dt = 0$. Since the complete elliptic integral of the 1st kind, $K(r/a)$, diverges to $+\infty$ at $r/a = 1$, the slip rate reverses in sign at the leading edge of the contact, *i.e.* at $(r, \theta) = (a, 0)$ and nearby. It is thus strictly inconsistent there with the assumed direction of the shear traction in the slipping contact circle. This situation is analogous to the issue explored by Adams *et al.* [10], who found a similar reversal of the slip rate for a sliding square bottomed punch. Note that slip reversal in Eq. (28) can only be avoided at the leading edge of the contact if $da/dt = -dD/dt$, in which case the leading edge is not progressing over the substrate.

We assume that, in the solutions we have used above, the same difficulty of slip rate reversal occurs at the leading edge of the sphere's contact circle for any value of Poisson's ratio, and for the case of limited slip as well as for gross slip.

We note that the situation just described strictly violates the 2nd Law of Thermodynamics locally, as work is being extracted from the interface where the slip rate reversal is occurring. However, for want of a more consistent solution we shall proceed, with the proviso that we should avoid violating the 2nd Law of Thermodynamics on a global basis and require that the total rate of heat generation must be no less than zero. It is anticipated that there will be situations where the inconsistency of slip rate reversal does not cause a big difference to the assessments subsequently made. Indeed, Adams *et al.* [10] comment that it is a reasonable engineering solution to ignore the slip rate reversal, and Hills, Sackfield and Churchman [11], in their discussion of Adams *et al.* [10], make similar points.

We return to the more general case with nonzero Poisson's ratio and observe that, upon calculation of Eq. (24), the terms in Eq. (27) dependent on $\partial\delta/\partial r$ and $\partial\delta/\partial\theta$ and those containing $\cos 2\theta$, due to symmetries in the functions involved, will self-cancel. Therefore, we compute Eq. (24) as

$$\frac{dQ}{dt} = \tau_o \int_A \left(\frac{\partial \tilde{\delta}}{\partial D}\frac{dD}{dt} + \frac{\partial \tilde{\delta}}{\partial a}\frac{da}{dt}\right) dA = \tau_o \int_A \left[\frac{dD}{dt} - \frac{2(2-\nu)\tau_o}{(1-\nu)\pi E^*}K\left(\frac{r}{a}\right)\frac{da}{dt}\right] dA$$
$$= \tau_o \pi a^2 \frac{dD}{dt} - \frac{4(2-\nu)\tau_o^2 a^2}{(1-\nu)E^*}\frac{da}{dt} \tag{29}$$

where $\tilde{\delta}$ is the slip given by Eq. (25) omitting the term containing $\cos 2\theta$. We substitute (29) into Eq. (22) with $\mathcal{E}_S = 0$ and obtain

$$\tau_o \pi a^2 \frac{dD}{dt} - \frac{4(2-\nu)\tau_o^2 a^2}{(1-\nu)E^*}\frac{da}{dt} = \tau_o \pi a^2 \frac{dD}{dt} - E^*\left(\frac{a^2}{R}-\Delta\right)^2 \frac{da}{dt} - \frac{4(2-\nu)\tau_o^2 a^2}{(1-\nu)E^*}\frac{da}{dt} + 2\pi a w_o \frac{da}{dt} \tag{30}$$

with $T$ having been evaluated through Eq. (16) and $d\mathcal{E}_D/dt$ from Eq. (21). The result is that

$$E^*\left(\frac{a^2}{R}-\Delta\right)^2 = 2\pi a w_o \tag{31}$$

consistent with Eq. (7). This confirms that energy balance plus the assumption that all slip work is dissipated as heat leads to the result that slip has no effect on adhesion. Furthermore, this outcome supports the step that was used to proceed from Eq. (22) to Eq. (11) in the case of limited slip.

We note that Eq. (31) is consistent with Eq. (4), and also with the reversible energy release rate, $\mathcal{G}^{re}$, being equal to $\mathcal{G}^{SE}$ as given in that equation. This indicates that, in the special case where slip has no influence on adhesion, the reversible energy release rate is also independent of terms such as $\mathcal{E}_D$ associated with the motion of the sphere parallel to the substrate. This observation disagrees with a result of Menga *et al.* [7], who find that, in gross slip, the reversible energy release rate, $\mathcal{G}^{re}$, has a contribution that arises from changes in $\mathcal{E}_D$. However, Menga *et al.* [7] withdrew this result in a corrigendum, and as result come into agreement with our Eq. (31). This implies that, as far as the corrigendum is concerned, they assume that all frictional slip is dissipated as heat.

*Case B: Slip has some effect on adhesion.*
Now consider the situation where some of the slip work is stored reversibly and the rest is dissipated as heat. At each instant we assume that the total friction work rate is the sum of a reversible part and an irreversible part. Thus, per unit area of the slipping interface, the total slip work rate is divided up according to

$$\tau_o \frac{d\delta}{dt} = \tau_o \frac{d\delta^{re}}{dt} + \tau_o \frac{d\delta^{irr}}{dt} \tag{32}$$

where $\delta^{re}$ is the reversible part of $\delta$ and $\delta^{irr}$ is the irreversible part. These add up to the total slip

$$\delta = \delta^{re} + \delta^{irr} \tag{33}$$

Since $\delta^{re}$ is reversible, the reversible shear work stored in the interface is

$$\mathcal{E}_S = \tau_o \int_A \delta^{re} dA \tag{34}$$

with the integrand being, of course, zero in the non-slipping segment of the interface within $0 \leq r \leq b$. Since the irreversible slip rate generates heat, we conclude that the rate of heat generation is

$$\frac{dQ}{dt} = \tau_o \int_A \frac{d\delta^{irr}}{dt} dA = \tau_o \int_A \left(\frac{d\delta}{dt} - \frac{d\delta^{re}}{dt}\right) dA \tag{35}$$

We first consider limited slip. In that case, $\mathcal{E}_D$ is given by Eq. (18) and the total slip by Eq. (23). Substitution of these into Eq. (22) provides

$$\frac{2\tau_o^2(2-v)}{\pi(1-v)E^*} \int_A \left\{\frac{d}{dt} \int_b^r \frac{\rho \cos^{-1}\frac{\rho}{r}}{\sqrt{a^2-\rho^2}} d\rho\right\} dA - \tau_o \int_A \frac{d\delta^{re}}{dt} dA$$
$$= \frac{1}{2} T \frac{dD}{dt} - \frac{1}{2} D \frac{dT}{dt} - E^* \left(\frac{a^2}{R} - \Delta\right)^2 \frac{da}{dt}$$
$$+ \frac{2(2-v)\tau_o^2}{(1-v)E^*} \frac{d}{dt} \int_b^a r \int_b^r \frac{\rho \cos^{-1}\frac{\rho}{r}}{\sqrt{a^2-\rho^2}} d\rho \, dr - \tau_o \frac{d}{dt} \int_A \delta^{re} dA + 2\pi a w_o \frac{da}{dt}$$
$$\tag{36}$$

where we have recognized that, after completion of the integrals, the term in Eq. (23) containing $\cos 2\theta$ will cancel. Also, we note that differentiation of $r$ with respect to time at a given material point in the substrate is omitted in the procedures to be used as the consequential result will also cancel upon completion of the integrals.

We note that Kim *et al.* [2] demonstrated that

$$\frac{d}{dt} \int_A \delta^{re} dA = \int_A \frac{d\delta^{re}}{dt} dA + \bar{\delta}_o^{re} \frac{dA}{dt} = \int_A \frac{d\delta^{re}}{dt} dA + 2\pi a \bar{\delta}_o^{re} \frac{da}{dt} \tag{37}$$

where $\bar{\delta}_o^{re}$ is the average of the reversible slip around the perimeter of contact. We thus obtain a simplification of Eq. (36) in the form

$$\frac{2(2-v)\tau_o^2}{(1-v)E^*} \int_b^a r \left\{ \frac{d}{dt} \int_b^r \frac{\rho \cos^{-1}\frac{\rho}{r}}{\sqrt{a^2-\rho^2}} d\rho \right\} dr$$

$$= \frac{1}{2} T \frac{dD}{dt} - \frac{1}{2} D \frac{dT}{dt} - E^* \left(\frac{a^2}{R} - \Delta\right)^2 \frac{da}{dt} - 2\pi a \tau_o \bar{\delta}_o^{re} \frac{da}{dt}$$

$$+ \frac{2(2-v)\tau_o^2 a}{(1-v)E^*} \int_b^a \frac{\rho \cos^{-1}\frac{\rho}{a}}{\sqrt{a^2-\rho^2}} d\rho \frac{da}{dt} + 2\pi a w_o \frac{da}{dt}$$

(38)

To obtain the next step forward we consider the case where $\delta^{re} = 0$, in which case Eq. (31) is valid, and obtain

$$\frac{2(2-v)\tau_o^2}{(1-v)E^*} \int_b^a r \left\{ \frac{d}{dt} \int_b^r \frac{\rho \cos^{-1}\frac{\rho}{r}}{\sqrt{a^2-\rho^2}} d\rho \right\} dr = \frac{1}{2} T \frac{dD}{dt} - \frac{1}{2} D \frac{dT}{dt} + \frac{2(2-v)\tau_o^2 a}{(1-v)E^*} \int_b^a \frac{\rho \cos^{-1}\frac{\rho}{a}}{\sqrt{a^2-\rho^2}} d\rho \frac{da}{dt} \quad (39)$$

thereby finessing the need to tackle some challenging integrals. It follows that Eq. (38) leads to

$$\frac{E^*}{2\pi a} \left(\frac{a^2}{R} - \Delta\right)^2 + \tau_o \bar{\delta}_o^{re} = w_o \tag{40}$$

demonstrating the reversible energy release rate, in this case, is

$$\mathcal{G}^{re} = \frac{E^*}{2\pi a} \left(\frac{a^2}{R} - \Delta\right)^2 + \tau_o \bar{\delta}_o^{re} \tag{41a}$$

Note the analogy to the Maugis-Dugdale model for adhesion [5] where, in the absence of slip, the reversible energy release rate is $\sigma_o H$ where $\sigma_o$ is the adhesive traction and $H$ is the gap between the 2 material surfaces at the perimeter of the contact.

We can also write the reversible energy release rate in terms of the applied load, in which case

$$\mathcal{G}^{re} = \frac{E^*}{2\pi a} \left(\frac{2a^2}{3R} - \frac{P}{2E^*a}\right)^2 + \tau_o \bar{\delta}_o^{re} \tag{41b}$$

Now consider gross slip, with $\mathcal{E}_D$ given by Eq. (21), $\mathcal{E}_S$ still given by Eq. (34) and slip by Eq. (25). Since Eq. (37) is still valid, Eq. (22) leads to

$$\tau_o \pi a^2 \frac{dD}{dt} - \frac{2(2-v)\tau_o^2}{(1-v)\pi E^*} \int_A \left\{ \frac{d}{dt} \left[aL\left(\frac{r}{a}\right)\right] \right\} dA$$

$$= T \frac{dD}{dt} - E^* \left(\frac{a^2}{R} - \Delta\right)^2 \frac{da}{dt} - \frac{4(2-v)\tau_o^2 a^2}{(1-v)E^*} \frac{da}{dt} - 2\pi a \tau_o \bar{\delta}_o^{re} \frac{da}{dt} + 2\pi a w_o \frac{da}{dt}$$

(42)

where, as previously, the term in Eq. (25) containing $\cos 2\theta$ and the differentiation with respect to time in the integral on the left hand side involves only $a$ and not $r$. We note that

$$\int_A \left\{ \frac{d}{dt}\left[ aL\left(\frac{r}{a}\right)\right]\right\} dA = 2\pi \int_0^a rK\left(\frac{r}{a}\right) dr \frac{da}{dt} = 2\pi a^2 \frac{da}{dt} \tag{43}$$

and Eq. (42) becomes Eq. (40), showing that Eq. (41) is still the reversible energy release rate.

We note that Eq. (41) is a special case of a more general result derived by Kim *et al.* [2], but it has now been obtained for the case where JKR adhesion prevails in the absence of slip. We note further, however, that the result in Eq. (41) is not useful until we have a model for $\delta^{re}$.

**Johnson's Model of Friction-Adhesion Interaction**
Stemming from concepts developed by Savkoor and Briggs [12], Johnson's [1] model of the effect of friction on the adhesion of a sphere to a flat substrate is in the form

$$w = w_o f\left(\frac{\tau_o \bar{\delta}_o}{\sigma_o h_o}\right) = \sigma_o h + \tau_o \bar{\delta}_o \tag{44}$$

where $w$ is the effective energy of adhesion during motion and loading of the sphere parallel to the substrate, $\bar{\delta}_o$ is the average of the contact perimeter frictional slip of the sphere relative to the substrate, and $f$ is a dimensionless function of its argument. The expressions in Eq. (44) are written in terms of Maugis-Dugdale [5] adhesion parameters where $w_o = \sigma_o h_o$ in which $\sigma_o$ is the Dugdale adhesive traction in the absence of friction, $h_o$ in the absence of friction is the interaction separation beyond which the Dugdale adhesive traction falls to zero, and $h$ is its value when there is frictional slip. Note that we have given a slightly different version of the model compared to the one first presented by Johnson [1], as he imagined that frictional slip would affect the Dugdale adhesive traction, but not the interaction distance. Instead, we have imagined here that frictional slip affects the interaction distance, but not the adhesive traction.

From the Savkoor [8] solution in Eq. (23), the frictional slip at the perimeter of the contact between the sphere and the substrate is

$$\delta_o = \frac{2\tau_o a}{\pi(1-\nu)E^*}\left[(2-\nu)\left(\sqrt{1-\frac{b^2}{a^2}}\cos^{-1}\frac{b}{a} + \frac{b}{a} - 1\right) - \frac{\nu}{3}\left(1-\frac{b^3}{a^3}\right)\cos 2\theta\right] \qquad D \leq \frac{(2-\nu)\tau_o a}{(1-\nu)E^*} \tag{45a}$$

for limited slip and from Eq. (25)

$$\delta_o = \frac{2\tau_o a}{\pi(1-\nu)E^*}\left[(2-\nu)\left(\frac{\pi}{2} - 1\right) - \frac{\nu}{3}\cos 2\theta\right] + D - \frac{(2-\nu)\tau_o a}{(1-\nu)E^*} \qquad D \geq \frac{(2-\nu)\tau_o a}{(1-\nu)E^*} \tag{45b}$$

for gross slip. Note that Eq. (45a) corrects an error in [1, 2] in the coefficient of $\cos 2\theta$, and a sign error in [1]. However, these errors are of no consequence as we require the average of the perimeter slip, given by

$$\bar{\delta}_o = \frac{2(2-\nu)\tau_o a}{\pi(1-\nu)E^*}\left(\sqrt{1-\frac{b^2}{a^2}}\cos^{-1}\frac{b}{a}+\frac{b}{a}-1\right) \qquad D \leq \frac{(2-\nu)\tau_o a}{(1-\nu)E^*} \qquad (46)$$

and

$$\bar{\delta}_o = D - \frac{2(2-\nu)\tau_o a}{\pi(1-\nu)E^*} \qquad D \geq \frac{(2-\nu)\tau_o a}{(1-\nu)E^*} \qquad (47)$$

Johnson's [1] interaction model was inspired by mixed-mode fracture mechanics, for which he referenced Hutchinson [13], though such ideas predated that paper. In mixed-mode fracture mechanics, the presence of shearing loads relative to the crack, by increasing the effect of dissipative processes during crack growth, can effectively increase the fracture toughness of the material, with the commonest assumption being that the fracture toughness is a function of the ratio of Mode II to Mode I stress intensity factors [13]. Johnson [1] used this analogy to justify Eq. (44), with $\tau_o \bar{\delta}_o$ playing the role of the Mode II stress intensity factor and, in our version of the model, $\sigma_o h_o$ that of the Mode I parameter. However, Kim *et al.* [2] later rationalized Johnson's [1] model. They showed that, in the Maugis-Dugdale model [5], the reversible energy release rate can only be identified in terms of parameters associated with the perimeter of the contact, and that only the reversible energy release rate can be equal to the adhesion energy, $w_o$. Recall that $w_o$ is the adhesion energy prevailing in the absence of frictional effects associated with slip parallel to the substrate. In other words, the correct interpretation of Johnson's [1] model is that it aims to identify accurately the reversible energy release rate so that the energy made available by it can supply the true, thermodynamically equilibriated adhesion energy, $w_o$. An important conclusion from this is that Johnson's [1] friction-adhesion interaction model, despite its inspiration from mixed-mode fracture mechanics, is not restricted to special cases of adhesion such as those where frictional slip is limited, but is quite versatile and can account for gross slip. Indeed, the form that Johnson [1] chose to present for the model is consistent with such applicability, as it is given in terms of Maugis-Dugdale [5] parameters. Thus, regimes other than JKR and limited slip are accommodated. In addition, the model should not be thought of simply as one dependent on the assumptions of the Maugis-Dugdale model [5]. For example, the parameters $\sigma_o$, $h$ and $h_o$ can be regarded as effective values characterizing other adhesion models, such as van der Waals attraction with a hard wall repulsion, or a Lennard-Jones interaction [14].

The rationalization of Johnson's [1] model given by Kim *et al.* [2] has the form

$$\mathcal{G}^{re} = \mathcal{G}^{re}(h, \bar{\delta}_o, \sigma_o, \tau_o) = w_o \qquad (48)$$

where $\mathcal{G}^{re}$ is the reversible energy release rate, with its functional form in Eq. (48) reflecting the fact that it can only be computed from parameters associated with the edge of the contact. Kim *et al.* [2] gave a rather technical presentation showing that other measures of the energy release rate can disagree with $\mathcal{G}^{re}$, and are then faulty in regard to the proper thermodynamic equilibrium for adhesion with frictional slip. We will not repeat the details here, and that paper can be consulted for them. From Eq. (48), and using dimensional analysis, Kim *et al.* [2] deduced that a friction/adhesion interaction model must have the form

$$F\left(\frac{\sigma_o h}{w_o}, \frac{\tau_o \bar{\delta}_o}{w_o}, \frac{\tau_o}{\sigma_o}\right) = 0 \tag{49}$$

where $F$ is a function of its arguments. The form of $F$ should be evaluated by experiment or, if possible, by theoretical computation. We note that the rightmost equality in Eq. (44) is consistent with Eq. (49), and in fact

$$F = f\left(\frac{\tau_o \bar{\delta}_o}{w_o}\right) - \frac{\sigma_o h}{w_o} - \frac{\tau_o \bar{\delta}_o}{w_o} \tag{50}$$

The interaction function chosen by Johnson [1] is

$$f(g) = \sqrt{(1+g)^2 - 2\alpha g} \tag{51a}$$

$$g = \frac{\tau_o \bar{\delta}_o}{w_o} \qquad \bar{\delta}_o \leq \bar{\delta}_o^m \tag{51b}$$

$$g = \frac{\tau_o \bar{\delta}_o^m}{w_o} = g_o \qquad \bar{\delta}_o \geq \bar{\delta}_o^m \tag{51c}$$

where Eq. (51c) corrects a misprint in [1], $\alpha$ is an interaction parameter and $\bar{\delta}_o^m$ is the maximum amount of slip that can affect adhesion. The interaction parameter is limited to

$$\alpha \leq \frac{(1+g_o)^2}{2g_o} \tag{52}$$

to ensure that the adhesion energy remains a real quantity.

When Eq. (51b) is valid, the interaction behavior is given by

$$\frac{\sigma_o h}{w_o} = \sqrt{\left(1 + \frac{\tau_o \bar{\delta}_o}{w_o}\right)^2 - 2\alpha \frac{\tau_o \bar{\delta}_o}{w_o}} - \frac{\tau_o \bar{\delta}_o}{w_o} \tag{53}$$

Thus, when $\alpha > 0$ frictional slip reduces the adhesion, whereas when $\alpha < 0$, adhesion is enhanced. With positive values of $\alpha$ approximately equal to 0.2, Johnson [1] found his model to be consistent with experiments then available. Furthermore, since Eq. (53) can be rearranged to provide

$$w_o = \sqrt{(\sigma_o h + \tau_o \bar{\delta}_o)^2 - \alpha(2-\alpha)(\tau_o \bar{\delta}_o)^2} - (1-\alpha)\tau_o \bar{\delta}_o \tag{54}$$

use of Eq. (48) tells us that Johnson's [1] friction/adhesion interaction model can now be seen to be one where the reversible energy release rate is being calculated as

$$\mathcal{G}^{re} = \sqrt{(\sigma_o h + \tau_o \bar{\delta}_o)^2 - \alpha(2-\alpha)(\tau_o \bar{\delta}_o)^2} - (1-\alpha)\tau_o \bar{\delta}_o \tag{55}$$

Thus, when $\alpha > 0$ the reversible energy release rate is increased above $\sigma_o h$, consistent with the observation above that then frictional slip reduces adhesion. The implication is that some of the energy release that is associated with $\tau_o \bar{\delta}_o$ contributes to the balance of energy associated with adhesion.

However, only in specific circumstances does all of the slip work, $\tau_o \bar{\delta}_o$, contribute to the reversible energy release rate. Only when $\tau_o \bar{\delta}_o = 2w_o(\alpha - 1)$ with $\alpha > 1$ do we find that $\mathcal{G}^{re} = \sigma_o h + \tau_o \bar{\delta}_o$ and, for this situation to be relevant, it would have to occur before $\bar{\delta}_o$ reaches $\bar{\delta}_o^m$. For example, subject to it being positive, if we have $\alpha = 1 + g_o/2$, the reversible energy release rate would be $\mathcal{G}^{re} = \sigma_o h_o$ upon initiation of slip, and would evolve monotonically with slip until equal to $\mathcal{G}^{re} = \sigma_o h + \tau_o \bar{\delta}_o^m$ at $\bar{\delta}_o = \bar{\delta}_o^m$. It then would remain unchanged in form at $\mathcal{G}^{re} = \sigma_o h + \tau_o \bar{\delta}_o^m$ for slip such that $\bar{\delta}_o > \bar{\delta}_o^m$, and would do so for both the limited slip and gross slip regimes. Note, however, that energy balance at all times requires $\mathcal{G}^{re} = w_o$, so that the numerical value of $\mathcal{G}^{re}$ remains constant even as its dependence on $\bar{\delta}_o$ evolves.

As noted above, Johnson [1] found that a good fit with experiments by Savkoor and Briggs [12] and Carpick *et al.* [15] was obtained with $\alpha \approx 0.2$. He also inferred that a good choice for $g_o$ would be around unity, with $\bar{\delta}_o^m$ approximately 0.2 nm. For the case of $\alpha = 0.2$

$$\mathcal{G}^{re} = \sqrt{(\sigma_o h + \tau_o \bar{\delta}_o)^2 - 0.36(\tau_o \bar{\delta}_o)^2} - 0.8\tau_o \bar{\delta}_o \qquad \bar{\delta}_o \leq \bar{\delta}_o^m \qquad (56a)$$

and, with $g_o = 1$,

$$\mathcal{G}^{re} = \sqrt{(\sigma_o h + w_o)^2 - 0.36 w_o^2} - 0.8 w_o \qquad \bar{\delta}_o \geq \bar{\delta}_o^m \qquad (56b)$$

Since $\mathcal{G}^{re} = w_o$, in the case where Eq. (56b) prevails $\sigma_o h \approx 0.9 w_o$, so that the effective adhesion energy is reduced by about 10% by frictional slip. This means also that frictional slip is absorbing about 10% of the adhesion energy at the leading, slipping edge of the contact and returning about 10% of it at the trailing edge, so that the loading normal to the substrate only has to balance the remaining 90% of it to be in adhesive equilibrium.

**Discussion**
We have deduced from our model that in the JKR regime of adhesion, the behavior is controlled by Eq. (41). Therefore, with equilibrium given by

$$\mathcal{G}^{re} = w_o \qquad (57)$$

and an effective energy of adhesion

$$w = w_o - \tau_o \delta^{re} \qquad (58)$$

the JKR equilibrium behavior is given by Eq. (7) & (8) with $w_o$ replaced by $w$.

We note here that $\delta^{re}$ may be positive, it may be zero or it may be negative. The simplest possibility is that $\delta^{re} = 0$, so that adhesion is unaffected by frictional slip. This would mean that, when released upon shrinkage of the contact area, strain energy stored in the substrate due to friction forces would be dissipated entirely in frictional slip.

If $\delta^{re} > 0$, then frictional slip reduces adhesion, as has been mentioned above. We imagine this happening by the process of slip taking the pristine surface of the substrate and, due to locally heterogeneous deformations on the scale of natural heterogeneities, injecting surface microstructures into the material. We postulate that such surface microstructures can reversibly store some elastic strain energy within them. The resulting phenomena will have the effect of raising the energy of the interface between the sphere and the substrate compared to the situation where, at the interface, such surface microstructures are absent. This picture therefore bases itself on the absence, or lower prevalence, of such surface microstructures when purely normal motion and loading of the sphere and the substrate are involved. As well as storing strain energy reversibly, such surface microstructures are likely to act as roughness keeping the 2 surfaces from intimate contact, and thus directly reducing the effectiveness of the adhesion interaction [16].

If $\delta^{re} < 0$, then frictional slip enhances adhesion. We imagine this process happening by the converse of the model that leads to $\delta^{re} > 0$. Imagine that the free surface of the substrate, when exposed to air, is rife with surface microstructures that store some elastic strain energy. Imagine also that slip can eliminate some of these surface microstructures, and therefore reduce the energy of the interface compared to the situation prior to the elimination of such surface microstructures. To complete the picture, we assume that purely normal loading and motion of the sphere and substrate cannot eliminate surface microstructures, or does so less effectively than frictional slip. It follows also that elimination of such surface microstructures allows a more intimate contact between the surfaces, and therefore directly makes the adhesion more effective.

We note that Johnson [1] has identified experiments where, as noted above, frictional slip reduces adhesion, indicating that, if our model is valid, for such cases $\delta^{re} > 0$. In addition, Menga *et al.* [7] cite some experiments in which frictional slip enhances adhesion, showing that, if our model is valid, in those cases $\delta^{re} < 0$. For example, Krick *et al.* [16] identify an increase in the contact area during sliding of nitrile rubber spheres on borosilicate float glass; we note, however, that Krick *et al.* [16] attribute the observed increase in adhesion to a viscoelastic effect. Nevertheless, there seems to be evidence that, if our model is valid, there are some systems that fit into the adhesion reducing regime and others that fit into the adhesion enhancing regime. However, we have not accounted for all possibilities of how the system can behave, such as rate effects that can influence the prevalence of bond breaking versus bond healing. As Meng *et al.* [7] point out, such effects can cause the velocity of sliding to be important and may influence whether $\delta^{re}$ is effectively positive or effectively negative in any given experiment. In addition, we have not considered fluctuations in the processes, whether temporal or spatial.

We note also that the expressions in Eq. (41) predict an effect on adhesion even when $w_o = 0$, given that $\delta^{re}$ can be nonzero in such circumstances. This would introduce adhesion in systems that are inherently non-adhesive. However, we argue that all interfaces between unlike materials are adhesive, even if that phenomenon is effectively very small compared to other attributes such as elastic strain energy of distortion. In addition, all materials have surface energy, so that even interfaces between components that are alike in material will exhibit adhesion, albeit very small. We note also that some experiments suggest that frictional sliding only eliminates adhesion but does not convert adhesion to repulsion [3, 12, 15, 16]. This indicates that $\tau_o \delta^{re} \leq w_o$, so that if $w_o$ is small then so is the maximum value of $\tau_o \delta^{re}$. Thus, we arrive at the conclusion that, in the unlikely event of a system truly being adhesionless with $w_o = 0$, we would expect that $\tau_o \delta^{re} = 0$, and so adhesion would continue to be absent in conditions of frictional slip.

We have noted above that Johnson's model [1] can be regarded as one where Eq. (55) is the formula used to compute the reversible energy release rate. In the Maugis-Dugdale regime of adhesion [5], this implies that

$$\tau_o \delta^{re} = \mathcal{G}^{re} - \sigma_o h = \sqrt{(\sigma_o h + \tau_o \bar{\delta}_o)^2 - \alpha(2-\alpha)(\tau_o \bar{\delta}_o)^2} - (1-\alpha)\tau_o \bar{\delta}_o - \sigma_o h \qquad (59)$$

giving us a model for the contribution of frictional slip to the reversible energy release rate. Although the model is not confined to $\tau_o \bar{\delta}_o / \sigma_o h \ll 1$, it is instructive to consider it in such circumstances. Expanding Eq. (59) for $\tau_o \bar{\delta}_o / \sigma_o h \ll 1$, we find that then

$$\delta^{re} = \alpha \bar{\delta}_o + \cdots \qquad (60)$$

Thus, in these conditions the reversible slip is a fixed fraction of the average perimeter slip and thus, for limited slip

$$\delta^{re} = \frac{2\alpha(2-\nu)\tau_o a}{\pi(1-\nu)E^*}\left(\sqrt{1-\frac{b^2}{a^2}}\cos^{-1}\frac{b}{a} + \frac{b}{a} - 1\right) \qquad D \leq \frac{(2-\nu)\tau_o a}{(1-\nu)E^*} \qquad (61)$$

with $D$ and $T$ given by Eq. (12) & (13) respectively. Written in terms of Eq. (41b) the total reversible energy release rate for the incompressible case then becomes

$$\mathcal{G}^{re} = \frac{E^*}{2\pi a}\left(\frac{2a^2}{3R} - \frac{P}{2E^*a}\right)^2 + \frac{6\alpha\tau_o^2}{\pi E^*}\left[\frac{E^*D}{3\tau_o}\cos^{-1}\sqrt{1-\left(\frac{E^*D}{3\tau_o a}\right)^2} + \sqrt{a^2 - \left(\frac{E^*D}{3\tau_o}\right)^2} - a\right] \qquad (62)$$

where we have rewritten $b$ in terms of $D$ via Eq. (12). When set equal to $w_o$, Eq. (62) becomes a rather complicated formula for the effect of frictional slip on the radius of contact, $a$, giving the latter as a function of the compressive load, $P$, and the motion, $D$, of the sphere parallel to the substrate.

To provide further insight, we consider the case of small displacement parallel to the substrate, i.e. $E^*D \ll 3\tau_o a$, in which case $\delta^{re} = \alpha E^* D^2/(3\pi\tau_o a) + \cdots$ and, to leading order, Eq. (62) becomes

$$\mathcal{G}^{re} = \frac{E^*}{2\pi a}\left(\frac{2a^2}{3R} - \frac{P}{2E^*a}\right)^2 + \frac{\alpha E^* D^2}{3\pi a} = w_o \qquad (63)$$

and simultaneously Eq. (41a) leads to

$$\mathcal{G}^{re} = \frac{E^*}{2\pi a}\left(\frac{a^2}{R} - \Delta\right)^2 + \frac{\alpha E^* D^2}{3\pi a} = w_o \tag{64}$$

We rearrange these in classic JKR form to read

$$P = \frac{4E^* a^3}{3R} - \sqrt{8\pi a^3 E^* w_o - \frac{8\alpha(E^*)^2 a^2 D^2}{3}} \tag{65}$$

$$\Delta = \frac{a^2}{R} - \sqrt{\frac{2\pi a w_o}{E^*} - \frac{2\alpha D^2}{3}} \tag{66}$$

We now consider the generation of heat to ensure that it will be positive. However, due to complexities in the analysis, we do this only for the special case where Eq. (60) is valid, in which case repetition of some of the previous algebra leads us to conclude that

$$\frac{dQ}{dt} = T\frac{dD}{dt} - \frac{d\varepsilon_D}{dt} - \alpha\tau_o \pi a^2 \frac{d\bar{\delta}_o}{dt} \tag{67}$$

For simplicity we address this result just after the transition from limited slip to gross slip on the assumption that then Eq. (60) still prevails. In that case, Eq. (67) leads to

$$\frac{dQ}{dt} = (1-\alpha)\tau_o \pi a^2 \frac{dD}{dt} - (2-\alpha)\frac{2(2-\nu)\tau_o^2 a^2}{(1-\nu)E^*}\frac{da}{dt} \tag{68}$$

When $\alpha = 0$, all frictional slip is dissipated as heat, and the result from Eq. (68) must then be positive. Since we have assumed that $\alpha \ll 1$, it follows that when $\alpha$ is not equal to zero the result in Eq. (68) will still be positive, as required by the 2nd Law of Thermodynamics.

We also comment on the unfortunate aspect of slip rate reversal that is a feature of the solutions developed by Keer and Goodman [6] and Savkoor [8]. This slip rate reversal that extracts energy from the interface is unsatisfactory. A better solution would not exhibit such a deficiency, but we are unaware of such an alternative. It is to be hoped that our insights are not held hostage by the deficient features of the solution. What can certainly be said is that the important results in the form of Eq. (41) are correct notwithstanding the slip rate reversal occurring in the Keer-Goodman-Savkoor [6, 8] solution, as they were originally derived by Kim *et al*. [2] independently of the details of the solutions in [6, 8]. Furthermore, it can also be said that an interface that converts all of the slip into heat will have $\delta^{re} = 0$ even after the slip rate reversal is eliminated as a feature of a solution. Thus, an interface that converts all slip into heat will have adhesion characteristics that are unaffected by slip. This conclusion is robust because there is no mechanism in such an interface to take shear displacements driven by strain energy release and convert them to adhesion or detachment energy.

**Conclusions**

We have considered the problem of an adherent rigid sphere sliding on a linear elastic, isotropic substrate in conditions where the radius of the contact is small compared to the radius of the sphere. Adhesion is assumed to obey the Johnson-Kendall-Roberts (JKR) model. We construct an energy balance, including the rate of dissipation of heat, the rate of external work, the rate of change of stored elastic strain energy, the rate of change of reversible energy stored in the interface, and the rate of change of adhesion energy. From this balance, we find that, when all frictional slip is dissipated as heat, adhesion is unaffected by slip. For a frictional model in which sliding is opposed by a uniform shear traction in the segment of the interface that slips, we find that the reversible energy release rate is that associated with JKR adhesion plus a contribution from slip. The latter is equal to the average of the reversible slip around the perimeter of the contact multiplied by the frictional shear traction. The balance of slip, the irreversible component, is dissipated as heat. We postulate that the reversible component of slip may be zero, in which case sliding has no influence on adhesion. We further conclude that if the reversible component of slip is nonzero, it may influence adhesion in a manner that can either increase it or decrease it. Through assessments of experiments made by Johnson, we find that in some cases the reversible component of slip is positive, and thus, in those cases, reduces the effectiveness of the adhesion. We point out that the approach described in this paper is a mixed-mode cohesive zone model equivalent to such formulations used in fracture mechanics, where John Hutchinson has performed much of the seminal research.

**Acknowledgment**

RMM acknowledges the support of an Alumni Award given by the Alexander von Humboldt Foundation, and a Leibniz Chair at the Leibniz Institute for New Materials, Saarbrücken, Germany. KSK acknowledges the support of NSF award CMMI-1563591.

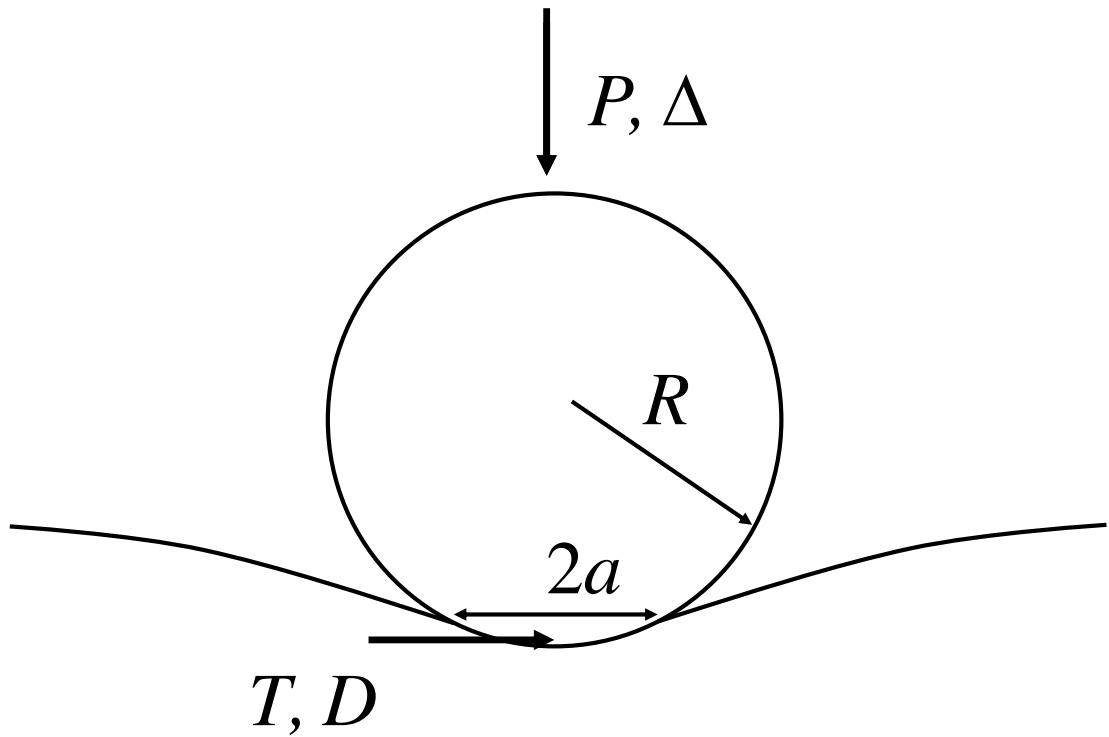

**Fig. 1**. A rigid sphere adhering to a linear elastic isotropic substrate and subject to applied loads $P$ and $T$ experiencing displacements $\Delta$ and $D$. The contact between the sphere and the substrate has radius $a$.

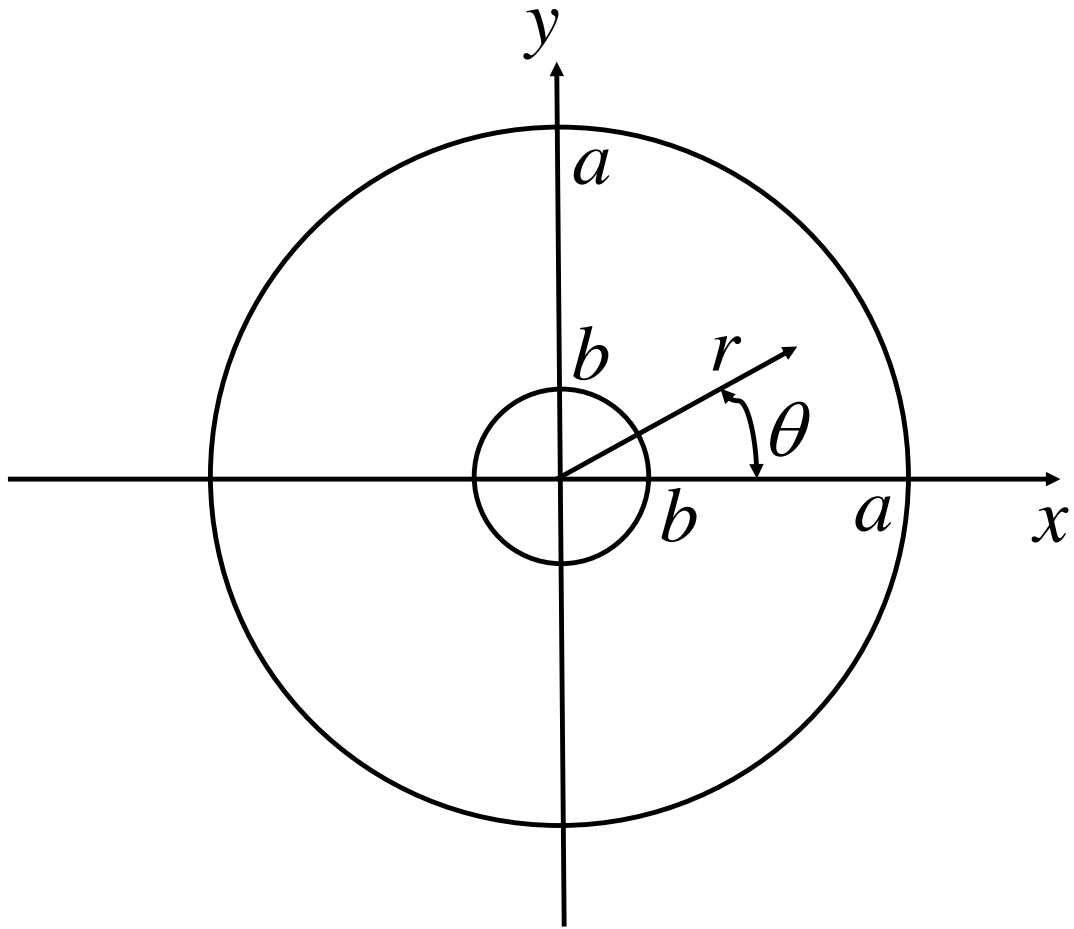

**Fig. 2.** The contact circle of radius $a$, showing the region that has not slipped within the circle of radius $b$. The polar coordinate system is also shown. In the slipping zone there is a uniform, constant traction of magnitude, $\tau_o$, opposing the motion $D$ of the sphere that is in the $x$ direction.